\documentclass[runningheads,a4paper]{llncs}
\usepackage{amssymb}

\usepackage{graphicx}

\usepackage{url}
\urldef{\mailsa}\path|{andriib,jad,loiret}@kth.se,omar.kacimi@gmail.com|
\newcommand{\keywords}[1]{\par\addvspace\baselineskip
\noindent\keywordname\enspace\ignorespaces#1}

\def\tightlist{}

\usepackage{lineno}
\usepackage{etoolbox}
\usepackage[usenames,dvipsnames,svgnames,table]{xcolor}
\usepackage{placeins}

\usepackage[capitalise]{cleveref}

\usepackage{tabularx}

\usepackage{wrapfig}

\usepackage[symbol]{footmisc}

\begin{document}

\mainmatter  % start of an individual contribution

\title{Improving lifecycle query in integrated toolchains using linked data and MQTT-based data warehousing\footnote[1]{The paper was presented at the ``IoT – Connected World and Semantic Interoperability Workshop'' on 22 October 2017 in Linz, Austria.}}

\titlerunning{Improving lifecycle query over integrated engineering toolchains}

\author{Andrii Berezovskyi \and Jad El-khoury \and Omar Kacimi \and Fr\'{e}d\'{e}ric Loiret}
\authorrunning{Improving lifecycle query over integrated engineering toolchains}

\institute{KTH Royal Institute of Technology,\\
Brinellv{\"a}gen 85, 100 44 Stockholm\\
\mailsa\\
\url{http://www.kth.se}}


\toctitle{Lecture Notes in Computer Science}
\tocauthor{Authors' Instructions}
\maketitle

\begin{abstract}

The development of increasingly complex IoT systems requires large engineering environments. These environments generally consist of tools from different vendors and are not necessarily integrated well with each other. In order to automate various analyses, queries across resources from multiple tools have to be executed in parallel to the engineering activities. In this paper, we identify the necessary requirements on such a query capability and evaluate different architectures according to these requirements.
We propose an improved lifecycle query architecture, which builds upon the existing Tracked Resource Set (TRS) protocol, and complements it with the MQTT messaging protocol in order to allow the data in the warehouse to be kept updated in real-time. As part of the case study focusing on the development of an IoT automated warehouse, this architecture was implemented for a toolchain integrated using RESTful microservices and linked data.

  \keywords{Internet of Things (IoT), tool integration, data warehousing, linked data, Resource Description Framework (RDF), Open Services for Lifecycle Collaboration (OSLC), Tracked Resource Set (TRS), SPARQL Protocol and RDF Query Language (SPARQL), Message Queue Telemetry Transport (MQTT)}
\end{abstract}

\section{Introduction}\label{introduction}

The explosive growth of IoT devices and systems is accompanied by their
increasing complexity. Modern engineering environments (toolchains)
required to develop such systems involve many tools and interdependent
processes, such as requirements analysis, embedded and cloud
architecture, prototyping, development, testing, deployment etc.

In order to facilitate these processes without impacting their quality
and flexibility, some integration has to take place to allow various
tools and processes to be connected. In systems engineering, one of the
important reasons to perform such integration is stimulated by the need
to perform a cross-system \emph{lifecycle query} (LCQ). LCQ allows one
to find resources that satisfy certain constraints defined across many
tools in the distributed toolchain. An example query that can be
relevant for a test engineer can be formulated in the following way:

\begin{quote}
\emph{Find all System Components that have been designed to satisfy the
Requirements with the status ``APPROVED'' and that have at least one
failed System Test, which has not been marked as ``RESOLVED''.}
\end{quote}

Another reason for building such a toolchain is that with the advent of
Big Data, the possibility to support and guide the systems engineering
efforts for the IoT systems by the use of the operational data becomes
viable. Such integration of operational data would also require advanced
query capabilities across raw data and over the insights generated by
applying various data science methods to the raw data.

Traditionally, a flavour of an extract-transform-load (ETL) process
would be used to transform and connect tool information. Analysed
against the 4Vs of the Big Data \cite{4vs}, one can argue that ETL
limits the velocity, variety and veracity of the data. Velocity is
reduced because the batch nature of ETL does not guarantee access to the
most recent data. Variety takes significant effort to achieve as each
integration requires an individual ETL pipeline. Finally, the
transformation step of each pipeline endangers veracity if semantics are
not preserved.

Instead, using microservices and Linked Data \cite{heath2011linked} to
integrate systems within the toolchain eliminates the need for multiple
transformations and ensures that the engineering environments of the IoT
systems are integrated correctly and efficiently. Assuming this approach
for the toolchain integration, one of the following architectures can be
used to add an LCQ capability to the toolchain:

\begin{enumerate}
\def\labelenumi{\arabic{enumi}.}
\tightlist
\item
  Directly query each microservice that contains the data needed to
  satisfy the query.
\item
  Rely on the distributed query support in the database systems, such as
  a SPARQL Federated Query \cite{sparql:federated}.
\item
  Build a Linked Data Warehouse (LDW) solution.
\end{enumerate}

This paper analyses the suitability of these approaches for IoT systems.
Based on this analysis, we identify their shortcomings when used in the
context of modern IoT systems. Instead, we propose an improved approach
where a data warehousing approach is complemented with a messaging
protocol to allow for a centralised LCQ capability while ensuring the
data is updated in real-time. This allows the LDW to run the queries
immediately after the engineering tool data has been updated.

We use a case study for the development of a robotic warehouse to
highlight the specifics of such architecture. The case study involves
tools for requirements analysis, change management as well as systems
engineering, which have to be integrated into a single toolchain.

In Section 2, we briefly present the background information on Linked
Data, MQTT, OSLC, and TRS. In Section 3, we present the case study. In
Section 4, we present the requirements for the LCQ architecture and
evaluate the available architectures according to them. In Section 5, we
propose an improved data warehousing architecture and detail its
implementation. Section 6 covers related work, followed by a discussion
in Section 7 and a conclusion in section 8.

\section{Background}\label{background}

Tool integration is performed between tools on different levels, such as
data format, user interface, reuse of functions
\cite{thomas1992definitions}. Using web-based Linked Data approach
allows the integrations to be made in a platform-independent and
scalable way \cite{mihindukulasooriya2013linked}. Resources are uniquely
identified using HTTP URIs and can link to other resources across such
integration, while representing the data using Resource Description
Framework (RDF) \cite{rdf11primer}.

Open Services for Lifecycle Collaboration (OSLC) is a set of
specifications that define requirements for the discovery capabilities
of the Linked Data services, their structure and \emph{resource shapes}.
This allows OSLC-compliant services to be used perform application
integration \cite{OSLCCore30,elaasar2013integrating}. OSLC builds on top
of the established web standards and best practices like HTTP, REST,
RDF, Linked Data Platform (LDP) \cite{ldp-spec}.

Using OSLC for building integration services allows defining domain
vocabularies separately from the services. These vocabularies can be
used for two main purposes: to communicate between services as well as
to translate between the tool-specific schema and the OSLC service
vocabulary. A common vocabulary for the services and their OSLC
capabilities can be defined using modelling tools
\cite{el2016modelling}, while the source code implementing the REST
services and OSLC resource classes can be generated from the model
\cite{el2016lyo}.

\section{Case Study}\label{case-study}

\begin{wrapfigure}{r}{0.5\textwidth}
  \begin{center}
    \vspace{-1.1cm}
    \includegraphics[width=0.5\textwidth,keepaspectratio]{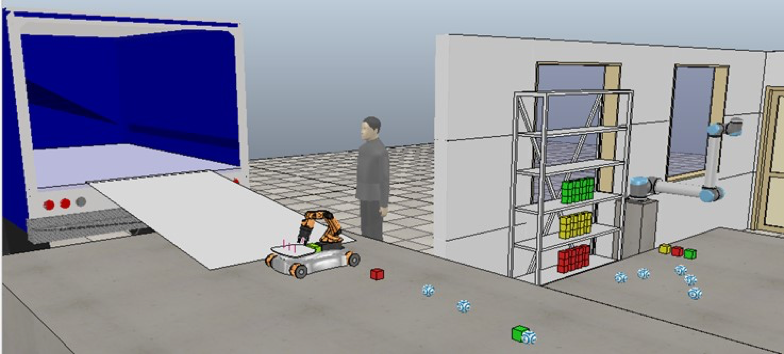}
    \vspace{-1.0cm}
  \end{center}
  \caption{\label{fig:scott}Robotic warehouse scenario.}
    \vspace{-1.1cm}
\end{wrapfigure}

Robotic warehouses are used to store and forward goods and require a
range of equipment (such as conveyor belts, identification and tagging
systems, self-driving vehicles and robots) in order to perform various
logistic tasks (\cref{fig:scott}). \emph{Logistics automation} is an
activity for optimising warehouse processes to improve its efficiency.

The efficiency of logistics in the robotic warehouse heavily depends on
the ``intelligence'' of a deployed IoT system. Therefore, the IoT
system's software needs to be continuously improved to optimise the
warehouse operations. Such improvements need to be supported by an
efficient development toolchain. The toolchain in the use-case
integrates the following tools:

\begin{enumerate}
\def\labelenumi{\arabic{enumi}.}
\tightlist
\item
  A Requirements Analysis tool to capture functional and non-functional
  requirements. In the use case, Eclipse ProR \cite{ProR} will be used
  for this purpose.
\item
  A Design tool to develop various parts of the IoT system. In the
  use-case, Matlab Simulink \cite{simulink} will be used to support
  model-based development approach.
\item
  A Change Management tool to ensure that when bugs arise or
  requirements change, the IoT system is changed a controlled manner. In
  the use-case, Bugzilla \cite{bugzilla} will be used for this purpose.
\end{enumerate}

Originally, these tools have no built-in integration between them, even
if they expose a certain API. However, entities in each of these tools
are naturally related to the entities in the other tools. With no
integration between them, such relationships are implicitly defined. For
example, a change request \textbf{CR1} in the Bugzilla tool may refer a
requirement \textbf{R1} in the text description, which does not allow to
define an explicit traceability link between the resources. As a
precondition to performing an accurate LCQ across the toolchain, an
\emph{integration capability} is needed to make those links explicit.

\begin{wrapfigure}{r}{0.6\linewidth}
  \begin{center}
    \vspace{-1.15cm}
    \includegraphics[width=\linewidth]{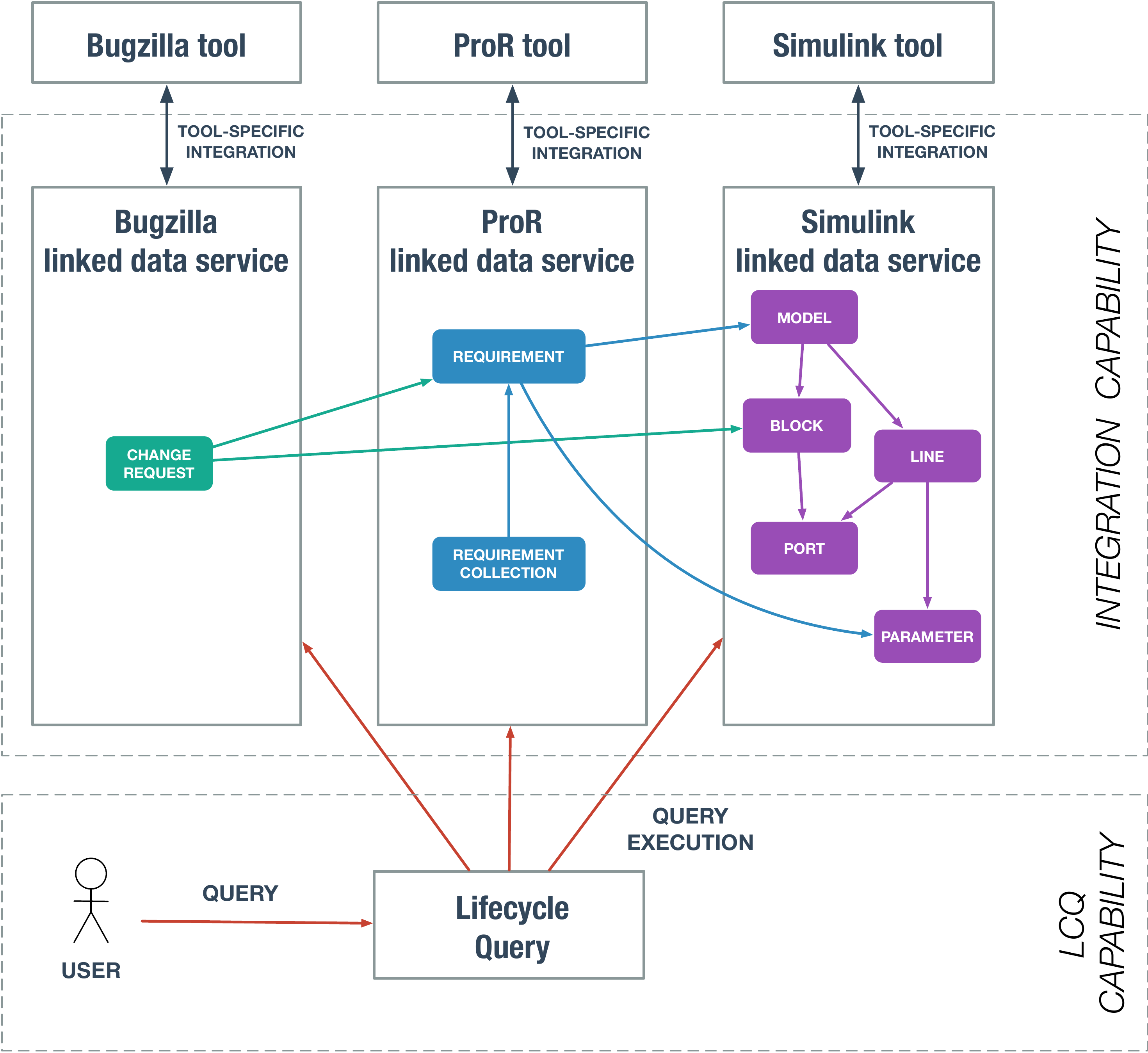}
    \vspace{-1cm}
  \end{center}
  \caption{\label{fig:toolchain}Overview of a Linked Data Toolchain, clarifying the distinction between its Integration (above) and Query (below) capabilities.}
  \vspace{-1.15cm}
\end{wrapfigure}

The Linked Data Toolchain presented in the use-case consists of three
logical parts (\cref{fig:toolchain}):

\begin{enumerate}
\def\labelenumi{\arabic{enumi}.}
\tightlist
\item
  A set of tools that make up a toolchain.
\item
  A set of corresponding Linked Data services, which provide an
  \emph{integration capability} for the toolchain.
\item
  A Lifecycle Query system, which provides an \emph{LCQ capability}.
\end{enumerate}

Linked Data services expose all entities managed by the underlying tools
as RDF resources via an OSLC-compliant RESTful service. The rest of the
paper assumes the that the integration capability exists and focuses on
the LCQ capability.

To evaluate different LCQ architectures, we define a number of analyses
that should be incorporated into the development process.\newline

\noindent
\textbf{LCQ1}\hspace{2mm} First, we want to ensure that every design
artefact was developed because a requirement demands it:

\begin{quote}
\emph{List all Simulink blocks that are not linked to any Requirement.}
\end{quote}

\noindent
\textbf{LCQ2}\hspace{2mm} Second, when considering a change request on a
requirement, a requirements analyst needs to assess the impact of a
requested change on other pending changes to the related requirements:

\begin{quote}
\emph{For a given Change Request \textbf{CR1} linked to the Requirement
\textbf{R1}, list all Change Requests \textbf{CRx} linked to the
Requirements \textbf{Rx} that refine the requirement \textbf{R1} as well
as all Change Requests \textbf{CRy} linked to the Requirements
\textbf{Ry} that are refined by the requirement \textbf{R1}.}
\end{quote}

\noindent
\textbf{LCQ3}\hspace{2mm} Finally, when a requirement or a design
artefact \emph{does not} change at all in a big project, it is also
suspicious and should be checked:

\begin{quote}
\emph{List all Models and Requirements which are not the subject of any
Change Request.}
\end{quote}

\section{Analysis of the lifecycle query
architectures}\label{analysis-of-the-lifecycle-query-architectures}

In this section, we analyse the LCQ capability needs for our case study,
to deduce and formalise the requirements they places on any LCQ
architecture (subsection \ref{ssec:lcq-req}). Afterwards, in the
subsection \ref{ssec:lcq-eval} we evaluate how different LCQ
architectures satisfy the defined requirements.

\subsection{Lifecycle query requirements}\label{ssec:lcq-req}

In the distributed Linked Data approach used for tool integration, the
lifecycle query capability has to satisfy a few constraints.\newline

\noindent
\textbf{RQ1}\hspace{2mm} In order to ensure that the LCQ results are not
obsolete by the time the queries are executed, LCQs should not run on
outdated data that differs significantly from the data operated by the
tools.

\begin{quote}
\emph{The lifecycle query system shall minimise the time difference
between the age of data used by the integration capability and the LCQ
capability.}
\end{quote}

\noindent
\textbf{RQ2}\hspace{2mm} One aspect that impacts the end-user usability
of an LCQ capability is the query performance itself. An LCQ that can
deliver real-time performance can be better integrated with the tool
interfaces, making it more valuable for engineers. A non-realtime but
relatively fast LCQ can be used for notifications. Finally, very slow
queries can only be delivered as reports.

\begin{quote}
\emph{The lifecycle queries should execute in minimum amount of time.}
\end{quote}

\noindent
\textbf{RQ3}\hspace{2mm} Sometimes, the execution of many queries can
introduce a lot of load on the system. LCQ capability should not
overload the integration capability.

\begin{quote}
\emph{The load the LCQ capability puts on the rest of the toolchain
shall be minimised.}
\end{quote}

\noindent
\textbf{RQ4}\hspace{2mm} Large development processes span a wide range
of tools, which can lead to a big toolchain. The LCQ capability shall be
able to scale accordingly.

\begin{quote}
\emph{The LCQ capability should efficiently scale to support larger
toolchains.}
\end{quote}

\noindent
\textbf{RQ5}\hspace{2mm} A toolchain configuration will change over
time, where tools may be added, removed or updated. For this reason, the
development of a new tool service to provide services to the LCQ
capability shall not be prohibitively complex.

\begin{quote}
\emph{Total effort needed to enable the LCQ capability for an individual
service should be minimised.}
\end{quote}

\noindent
\textbf{RQ6}\hspace{2mm} Changes in business requirements, development
processes, and engineering tools may introduce a need for new or
modified queries to collect new metrics or verify a certain state of
resources. The LCQ capability should be flexible enough to easily adapt
to the new requirements.

\begin{quote}
\emph{Effort required to add a new LCQ or modify an existing one shall
be minimised.}
\end{quote}

\noindent
\textbf{RQ7}\hspace{2mm} Toolchains include legacy tools and
implementing correct integrations requires a lot of effort. An
integration capability based on Linked Data also provides an additional
benefit of being tool- and implementation-independent. The LCQ
capability should fit into the toolchain well and should not nullify
those efforts.

\begin{quote}
\emph{The LCQ capability should require minimal architectural changes to
the toolchain and its integration based on Linked Data.}
\end{quote}

\subsection{Lifecycle query architecture
comparison}\label{ssec:lcq-eval}

Given the toolchain architecture, with an integration capability based
on Linked Data, the following ways can be used to implement the LCQ
capability:

\begin{enumerate}
\def\labelenumi{\arabic{enumi}.}
\tightlist
\item
  \emph{Direct Query} over REST, where the data is gathered manually
  through a series of HTTP requests to the respective services of each
  tool that holds the data to be queried.
\item
  \emph{Distributed Query} using the SPARQL Federated Query.
\item
  \emph{Data Warehousing} solution with a centralised query capability.
\end{enumerate}

While the Direct Query gives the developer most control over querying
and ensures that the data is always up-to-date, the developer must
determine the request order on a case by case basis (\emph{query
planning}) and make sure that the data is cached properly, in order to
avoid overloading the integration capability. Even with all these
concerns handled, developing new queries is non-trivial, whether the
developer decides to develop the queries programmatically or introduce a
simplified domain-specific language (DSL).

Switching to the Distributed Query, one can rely on the query planner of
a triplestore. The approach, however, has its own limitations:

\begin{itemize}
\tightlist
\item
  The triplestore of a specific service may cache responses from the
  certain calls but cannot determine the caching validity of the
  underlying RDF resources (as opposed to the manual query that can and
  should rely on the \texttt{Cache-Control} HTTP header for its REST
  calls). In particular, the SPARQL Federated Query specification does
  not cover caching and only allows to silently ignore errors to prevent
  cascading failure of a query \cite{sparql:federated}.
\item
  Similar to Direct Query, Distributed Query causes an increased load on
  the toolchain. The load has now instead shifted from the microservices
  to their triplestores. The expressiveness of SPARQL can allow an even
  higher load on the toolchain, which can lead to the SPARQL endpoint
  downtime \cite{daverog}.
\item
  Exposing a database is an anti-pattern, creating problems for access
  control \cite{costabello2013access} as well as limiting the use of
  patterns that rely on reified statements and named graphs
  \cite{linkedpatterns12}.
\item
  Writing Federated Queries is more cumbersome than plain SPARQL and
  expects the developer to statically define the query endpoints (using
  the \texttt{SERVICE} keyword). In most cases, the UI would have to be
  developed to abstract the query formulation from SPARQL.
\end{itemize}

Finally, the Data Warehousing appro-ach allows queries to be developed
using plain SPARQL without a concern for caching, error handling, or the
distributed nature of the tool-chain (including the risk of overloading
the tool-chain services with ``heavy'' queries). The main challenge is
to keep the data in the LDW updated with every service in the toolchain.

\setlength{\columnsep}{0.4cm}

\begin{wraptable}{r}{0.5\linewidth}
  \vspace{-15mm}
  \begin{center}
\caption{\label{tbl:lq-approaches} Comparison of the lifecycle query system architectures}
{\footnotesize
\begin{tabularx}{\linewidth}{ | X | c | c | c | }
\hline
& \rotatebox{90}{\parbox[t]{1cm}{Direct\\Query}} & \rotatebox{90}{\parbox[t]{1.75cm}{Distributed\\Query}}  & \rotatebox{89}{\parbox[t]{1.85cm}{Data\\Warehousing}} \\
 \hline
 RQ1 (min. data age) & ++ & ++ & -- \\
 \hline
 RQ2 (fast execution) & -- & + & ++  \\
 \hline
 RQ3 (don't overload the toolchain) & -- & + & ++  \\
 \hline
 RQ4 (scale) & + & -- & ++  \\
 \hline
 RQ5 (easy LCQ integration in each service) & ++ & + & -- \\
 \hline
 RQ6 (queries shall be easy to develop) & -- & + & ++  \\
 \hline
 RQ7 (min. architectural changes) & ++ & -- & +  \\
 \hline
\end{tabularx}}
  \vspace{-17mm}
  \end{center}
\end{wraptable}

\Cref{tbl:lq-approaches} summarises how different approaches satisfy the
original requirements of the lifecycle query. It uses the following
comparative values:

\begin{itemize}
\tightlist
\item
  ``++'' denotes that the requirement is fully satisfied
\item
  ``+'' denotes that the requirement is partly satisfied
\item
  ``--'' denotes that the requirement is poorly satisfied, additional
  work may be necessary to overcome the shortcoming
\end{itemize}

Each architecture has its own strengths and weaknesses, but in general,
Data Warehousing architecture has the best fit for the LCQ system. Data
Warehousing architecture has, however, two important issues:

\begin{enumerate}
\def\labelenumi{\arabic{enumi}.}
\tightlist
\item
  In most warehousing solutions, data in the LDW is updated through
  regular pull requests for changes to each of the tool services.
  Depending on the frequency of these updates, a difference in the data
  state will exist between the LDW data and the original data in the
  toolchain.
\item
  Integrating a new tool in the tool chain requires additional
  implementation in the corresponding Linked Data service to support LDW
  requests.
\end{enumerate}

While acknowledging the suitability of the data warehousing approach, we
present in the next section an improved architecture that addresses
these identified issues.

\section{Architecture}\label{architecture}

\begin{wrapfigure}{r}{0.65\linewidth}
  \begin{center}
    \vspace{-1.4cm}
  \includegraphics[width=\linewidth,keepaspectratio]{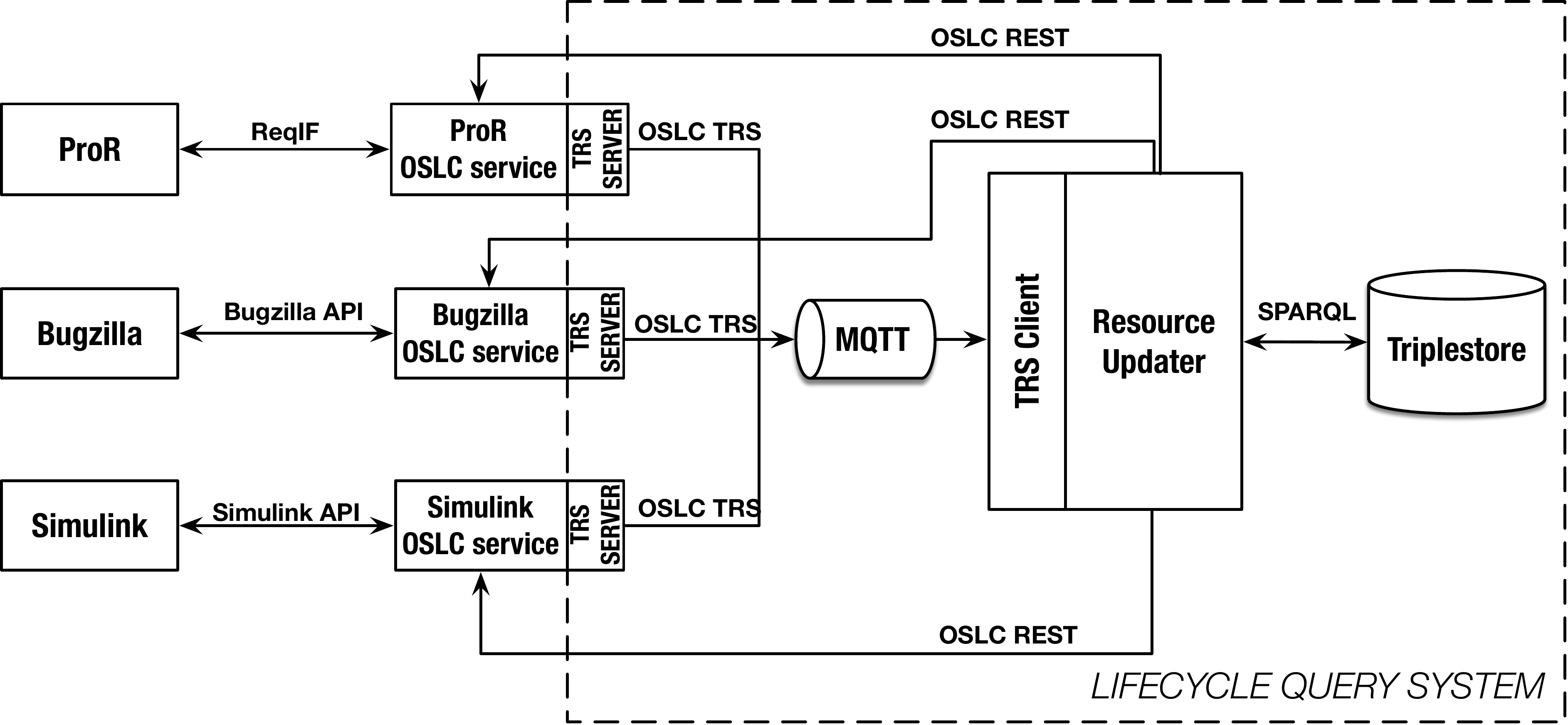}
    \vspace{-1cm}
  \end{center}
  \caption{\label{fig:arch-trs} The LCQ system architecture.}
  \vspace{-1.4cm}
\end{wrapfigure}

As illustrated in \cref{fig:arch-trs}, the LCQ system consists of 2
parts:

\begin{itemize}
\tightlist
\item
  the Tracked Resource Set (TRS) Client, and
\item
  the Resource Updater.
\end{itemize}

The \textbf{TRS client} is the core component of the LDW solution. It
implements the TRS protocol, which allows a server to expose resources
so that these changes can be discovered and tracked by the clients over
HTTP. Its main responsibilities are:

\begin{itemize}
\tightlist
\item
  initiating the synchronisation,
\item
  fetching all subsequent resource update events,
\item
  compacting the update event list in order to avoid applying multiple
  changes on the same resource (ie if the resource was created,
  modified, and deleted, nothing will be done at all).
\end{itemize}

Now, the TRS protocol implies a pull-based periodic approach to fetch
updates from the servers. The update period is defined by the TRS client
and can be adjusted per TRS server, depending on the frequency of
changes to the underlying resources. This, nevertheless, leads to a
delay between the source data and the information in the TRS client. To
remedy the staleness issue identified in the previous section, we
complement the TRS protocol with MQTT to eliminate the polling period
and shorten the LDW data update delay.

MQTT is a protocol to allow \emph{publish-subscribe}-based messaging
\cite{standard2014mqtt}. A \emph{message broker} is used to receive,
store, and distribute messages between clients. The small footprint of
the protocol made it attractive for the use in IoT systems, particularly
gateways \cite{mqtt-aws,mqtt-azure,mqtt-bluemix}.

Upon a change in a tool, the TRS server sends a \texttt{ChangeEvent}
message via MQTT containing 2 properties of a single change event (event
sequence number and the URI of the changed resource) and its type (one
of \texttt{Creation}, \texttt{Modification}, \texttt{Deletion}, as
defined in the TRS specification). The \textbf{Resource Updater} then
updates the RDF resources in the triplestore after reconstructing the
TRS Change Event from the \texttt{ChangeEvent} message.

In this architecture, any LCQ queries can be performed across the entire
dataset of the toolchain using SPARQL queries on a single endpoint.

\subsection{Implementation}\label{implementation}

The LCQ architecture presented in this paper was implemented and
integrated with the toolchain used in the case study. The implementation
consists of the lifecycle queries, which are available on
Github\footnote{https://github.com/berezovskyi/trs-mqhouse}, and 2
open-source modules, which have been contributed to the Eclipse Lyo
project\footnote{https://bugs.eclipse.org/bugs/show\_bug.cgi?id=513207}:

\begin{itemize}
\tightlist
\item
  The LDW service has been contributed as a TRS Consumer.
\item
  The implementation of a TRS Server has been extracted into a library
  that supports the REST API needed communication with the LDW in a
  JAX-RS service.
\end{itemize}

\subsection{Analysis}\label{analysis}

In order to assess how the proposed architecture compares to the
existing ones, we will reevaluate how it satisfies the original
requirements.

\textbf{RQ1}\hspace{2mm} The proposed architecture has still greater
latency than the Direct Query (DRQ) and Distributed Query (DSQ)
approaches but improves significantly against the Data Warehousing (DWH)
architecture without messaging. An important note here is that the DRQ
and DSQ approaches cannot achieve zero overhead because they communicate
with the linked-data--based OSLC service instead of directly
communicating with the tool. Depending on the tool integration method,
the difference between the proposed architecture and the DRQ/DSQ
approaches might be minimal.

\textbf{RQ2}\hspace{2mm} The proposed architecture makes no improvements
compared to the base DWH approach but takes advantage of its centralised
query capability.

\textbf{RQ3}\hspace{2mm} The proposed architecture fetches each resource
state at most once (some retrievals can be avoided due to change log
compaction) and is therefore optimal in this regard. However, the big
volume of small updates can increase the system load. For such cases,
both the TRS Client and the TRS Server library can batch a certain
number updates from a service.

\textbf{RQ4}\hspace{2mm} The publish-subscribe paradigm used in the
proposed architecture is inherently more suitable for building scalable
distributed systems \cite{eugster2003many}.

\textbf{RQ5}\hspace{2mm} The TRS Server library drastically reduces the
number of source lines of code (SLOC) needed for the implementation of a
compatible interface. It requires more effort than DRQ but less that
DSQ, especially if the OSLC service is not using a triplestore as an
underlying data source, but a database like MongoDB or simply caches the
Linked Data converted from a tool in a key-value storage like Redis.

\textbf{RQ6}\hspace{2mm} The DWH approach provides an ability to write
simple SPARQL queries and our proposed architecture does not affect
that.

\textbf{RQ7}\hspace{2mm} The proposed approach does not affect any of
the existing Linked Data architecture but introduces a messaging system,
which might exist in many enterprise integration platforms already
\cite{hohpe2003enterprise}. The problem arises when the source of the
OSLC service cannot be modified (especially when a proprietary tool
comes with a ready OSLC implementation, such as IBM DOORS NG
\cite{doorsng:oslc}). To that extent, our architecture allows those
tools to be integrated into the LDW using an unmodified TRS protocol.

\subsection{Future work}\label{future-work}

First, we would like to evaluate the proposed architecture on a
toolchain with a production load and identify ways to improve its
efficiency further. Potential solutions to improve performance include:
sending updated resource contents in a message together with a change
event, publishing the base set of resources in a compressed binary HDT
format \cite{fernandez2013binary}. Additionally, an OSLC service can
modified to stream change events immediately when the resource updates
are received from the underlying tool.

Finally, we would like to explore an architecture where each OSLC
service has both a TRS server and a TRS client components. This would
allow the resource changes to be distributed across all services in the
toolchain, not only to the TRS client of an LCQ system, which would make
the toolchain \emph{reactive}.

\section{Related work}\label{related-work}

Efforts to enable query over distributed data sources predate Linked
Data. Sheth and Larson have provided a taxonomy of multi-database
systems and relevant concepts \cite{sheth1990federated}. Levy et al.
\cite{Levy96} proposed the generation of query plans from complex
queries that detail the individual queries to be executed against the
data sources and the rules for combining the results of such distributed
query.

Support for distributed queries over RDF resources has been implemented
in the SPARQL 1.1 Federated Query \cite{harris2013sparql}, but the query
developer is responsible for specifying the subqueries against other
triplestores using the \texttt{SERVICE} keyword. Quilitz et al.
introduce DARQ, an engine that allows hiding these subqueries by
equipping the triplestore with the service description that is used for
query rewriting \cite{quilitz2008querying}. DARQ has serious limitations
and has been discontinued \cite{DARQFede98:online}.

SDShare is a protocol allows RDF resources to be synchronised between
servers \cite{SDSHARE}. The changes are published using Atom feeds
\cite{nottingham2005atom}. A collection of resources is described by a
feed containing a \emph{snapshot} of resources at a given time and a
feed containing the updated \emph{fragments} of the snapshot. Compared
to TRS, it lacks a concept of a \emph{cutoff event} that allows
rebuilding the TRS Base as opposed to the fixed SHShare Snapshot, among
other things.

In an effort to make SPARQL scalable and highly-available, Linked Data
fragments (LDF) have very promising results and represent a viable
approach to ``cloud-scale'' SPARQL
\cite{verborgh2014web,verborgh2014querying}.

\section{Conclusion}\label{conclusion}

In this paper, we presented the need for the LCQ capability in
integrated engineering environments. We also presented a case study
involving a toolchain integrated using OSLC-compliant microservices and
the requirements that such toolchain puts on the LCQ capability. Three
LCQ architectures were presented and evaluated according to the
identified requirements.

An improved LCQ architecture was proposed, which complements the TRS
protocol by delivering change events in messages over the MQTT protocol.
This allows reducing the delay between the original resource update and
the update of a corresponding resource in the LDW. The proposed
architecture has been evaluated according to the requirements and
compared to the previously discussed ones. The architecture provides
advantages over the plain LDW as well as over the Direct Query and the
Distributed Query architecture.

\paragraph{\textbf{Acknowledgement}}

This work is supported by the SCOTT project\footnote{\emph{SCOTT
  (www.scott-project.eu) has received funding from the Electronic
  Component Systems for European Leadership Joint Undertaking under
  grant agreement No~737422. This Joint Undertaking receives support
  from the European Union's Horizon 2020 research and innovation
  programme and Austria, Spain, Finland, Ireland, Sweden, Germany,
  Poland, Portugal, Netherlands, Belgium, Norway.}}.

% \subsection*{Acknowledgements}

\bibliographystyle{splncs03}
\bibliography{references}

\begin{thebibliography}{10}
\providecommand{\url}[1]{\texttt{#1}}
\providecommand{\urlprefix}{URL }

\bibitem{doorsng:oslc}
Access oslc services from ibm rational doors.
  \url{https://www.ibm.com/developerworks/rational/library/oslc-services-rational-doors/index.html},
  (Accessed on 2017-08-20)

\bibitem{mqtt-aws}
Aws iot platform - amazon web services.
  \url{https://aws.amazon.com/iot-platform/}, (Accessed on 2017-08-18)

\bibitem{mqtt-azure}
Azure iot suite | microsoft azure.
  \url{https://azure.microsoft.com/en-us/suites/iot-suite/}, (Accessed on
  2017-08-18)

\bibitem{bugzilla}
Bugzilla. \url{https://www.bugzilla.org/}, (Accessed on 2017-08-18)

\bibitem{DARQFede98:online}
{DARQ - federated queries with SPARQL}.
  \url{http://darq.sourceforge.net/#Limitations_and_known_issues}, (Accessed on
  2017-08-29)

\bibitem{daverog}
The enduring myth of the sparql endpoint | dave's blog.
  \url{https://daverog.wordpress.com/2013/06/04/the-enduring-myth-of-the-sparql-endpoint/},
  (Accessed on 2017-08-19)

\bibitem{mqtt-bluemix}
Ibm bluemix - internet of things.
  \url{https://www.ibm.com/cloud-computing/bluemix/internet-of-things},
  (Accessed on 2017-08-18)

\bibitem{4vs}
{Infographic: The Four {V}'s of Big Data | IBM Big Data \& Analytics Hub}.
  \url{http://www.ibmbigdatahub.com/infographic/four-vs-big-data}, (Accessed on
  2017-08-17)

\bibitem{ldp-spec}
Linked data platform 1.0. \url{https://www.w3.org/TR/ldp/}, (Accessed on
  2017-08-18)

\bibitem{OSLCCore30}
Oslc core version 3.0.
  \url{https://tools.oasis-open.org/version-control/svn/oslc-core/trunk/specs/oslc-core.html},
  (Accessed on 2017-08-18)

\bibitem{ProR}
{ProR Requirements Engineering Platform (part of Eclipse RMF)}.
  \url{https://www.eclipse.org/rmf/pror/}, (Accessed on 2017-08-18)

\bibitem{rdf11primer}
Rdf 1.1 primer. \url{https://www.w3.org/TR/2014/NOTE-rdf11-primer-20140225/},
  (Accessed on 2017-08-18)

\bibitem{simulink}
{Simulink - Simulation and Model-Based Design}.
  \url{https://se.mathworks.com/products/simulink.html}, (Accessed on
  2017-08-18)

\bibitem{sparql:federated}
{SPARQL 1.1 Federated Query}.
  \url{https://www.w3.org/TR/sparql11-federated-query/}, (Accessed on
  2017-08-17)

\bibitem{costabello2013access}
Costabello, L., Villata, S., Rocha, O.R., Gandon, F.: Access control for http
  operations on linked data. In: Extended Semantic Web Conference. pp.
  185--199. Springer (2013)

\bibitem{linkedpatterns12}
Dodds, L., Davis, I.: {Linked Data Patterns}.
  \url{http://patterns.dataincubator.org/book/data-management-patterns.html}
  (2012), (Accessed on 2017-08-19)

\bibitem{el2016lyo}
El-khoury, J.: Lyo code generator: A model-based code generator for the
  development of oslc-compliant tool interfaces. SoftwareX  5,  190--194 (2016)

\bibitem{el2016modelling}
El-Khoury, J., Gurdur, D., Loiret, F., T{\"o}rngren, M., Zhang, D., Nyberg, M.:
  Modelling support for a linked data approach to tool interoperability.
  ALLDATA 2016 p.~51 (2016)

\bibitem{elaasar2013integrating}
Elaasar, M., Neal, A.: Integrating modeling tools in the development lifecycle
  with oslc: A case study. In: International Conference on Model Driven
  Engineering Languages and Systems. pp. 154--169. Springer (2013)

\bibitem{eugster2003many}
Eugster, P.T., Felber, P.A., Guerraoui, R., Kermarrec, A.M.: The many faces of
  publish/subscribe. ACM computing surveys (CSUR)  35(2),  114--131 (2003)

\bibitem{fernandez2013binary}
Fern{\'a}ndez, J.D., Mart{\'\i}nez-Prieto, M.A., Guti{\'e}rrez, C., Polleres,
  A., Arias, M.: Binary rdf representation for publication and exchange (hdt).
  Web Semantics: Science, Services and Agents on the World Wide Web  19,
  22--41 (2013)

\bibitem{harris2013sparql}
Harris, S., Seaborne, A., Prud'hommeaux, E.: Sparql 1.1 query language. W3C
  recommendation  21(10) (2013)

\bibitem{heath2011linked}
Heath, T., Bizer, C.: Linked data: Evolving the web into a global data space.
  Synthesis lectures on the semantic web: theory and technology  1(1),  1--136
  (2011)

\bibitem{hohpe2003enterprise}
Hohpe, G., Woolf, B., et~al.: Enterprise integration patterns  (2003)

\bibitem{Levy96}
Levy, A., Rajaraman, A., Ordille, J.: Querying heterogeneous information
  sources using source descriptions. Tech. rep., Stanford InfoLab (1996)

\bibitem{mihindukulasooriya2013linked}
Mihindukulasooriya, N., Garc{\'\i}a-Castro, R., Esteban-Guti{\'e}rrez, M.:
  Linked data platform as a novel approach for enterprise application
  integration. In: Proceedings of the Fourth International Conference on
  Consuming Linked Data-Volume 1034. pp. 146--157. CEUR-WS. org (2013)

\bibitem{SDSHARE}
Moore, G., Garshol, L.M.: {SDShare} - a protocol for the syndication of
  resource descriptions. \url{http://www.sdshare.org/} (July 2012)

\bibitem{nottingham2005atom}
Nottingham, M., Sayre, R.: {RFC  4287}, the {Atom} syndication format  (2005)

\bibitem{quilitz2008querying}
Quilitz, B., Leser, U.: Querying distributed rdf data sources with sparql. In:
  European Semantic Web Conference. pp. 524--538. Springer (2008)

\bibitem{sheth1990federated}
Sheth, A.P., Larson, J.A.: Federated database systems for managing distributed,
  heterogeneous, and autonomous databases. ACM Computing Surveys (CSUR)  22(3),
   183--236 (1990)

\bibitem{standard2014mqtt}
Standard, O.: Mqtt version 3.1.  (2014)

\bibitem{thomas1992definitions}
Thomas, I., Nejmeh, B.A.: Definitions of tool integration for environments.
  IEEE software  9(2),  29--35 (1992)

\bibitem{verborgh2014querying}
Verborgh, R., Hartig, O., De~Meester, B., Haesendonck, G., De~Vocht, L.,
  Vander~Sande, M., Cyganiak, R., Colpaert, P., Mannens, E., Van~de Walle, R.:
  Querying datasets on the web with high availability. In: International
  Semantic Web Conference. pp. 180--196. Springer, Cham (2014)

\bibitem{verborgh2014web}
Verborgh, R., Vander~Sande, M., Colpaert, P., Coppens, S., Mannens, E., Van~de
  Walle, R.: Web-scale querying through linked data fragments. In: LDOW (2014)

\end{thebibliography}

\end{document}